\documentclass[11pt,onecolumn,journal]{IEEEtran}
\usepackage{amsmath,amsfonts}
\usepackage{algorithmic}
\usepackage{algorithm}
\usepackage{array}
\usepackage[caption=false,font=normalsize,labelfont=sf,textfont=sf]{subfig}
\usepackage{textcomp}
\usepackage{stfloats}
\usepackage{url}
\usepackage{verbatim}
\usepackage{graphicx}
\usepackage{cite}
\usepackage{hyperref}
\usepackage{booktabs}
\usepackage{xcolor}

\usepackage{units}
\usepackage{mathtools}

\usepackage{bm}

\usepackage{amssymb}
\usepackage{multirow}
\usepackage{url}

\begin{document}
%
\title{The IEEE Signal Processing Society's Leading Role in Developing Standards for Computational Imaging and Sensing: Part II}
%
\author{Andreas Bathelt,~\emph{Fraunhofer Institute for High Frequency Physics and Radar Techniques}, Benjamin Deutschmann,~\emph{Graz University of Technology}, Hyeon Seok Rou,~\emph{Constructor University}, Kuranage Roche Rayan Ranasinghe,~\emph{Constructor University}, Giuseppe Thadeu Freitas de Abreu,~\emph{Constructor University}, Peter Vouras,~\emph{U.S. Department of Defense}%

}
%
\markboth{IEEE Signal Processing Magazine,~Vol.~XX, No.~XX, June~2025}%
{Bathelt \MakeLowercase{\textit{et al.}}: The IEEE Signal Processing Society's Leading Role in Developing Standards for Computational Imaging and Sensing: Part II}


\maketitle

\begin{abstract}
In every imaging or sensing application, the physical hardware creates constraints that must be overcome or they limit system performance.  Techniques that leverage additional degrees of freedom can effectively extend performance beyond the inherent physical capabilities of the hardware.  An example includes synchronizing distributed sensors so as to synthesize a larger aperture for remote sensing applications.  An additional example is integrating the communication and sensing functions in a wireless system through the clever design of waveforms and optimized resource management.  As these technologies mature beyond the conceptual and prototype phase they will ultimately transition to the commercial market.  Here, standards play a critical role in ensuring success.  Standards ensure interoperability between systems manufactured by different vendors and define industry best practices for vendors and customers alike.  The Signal Processing Society of the Institute for Electrical and Electronics Engineers (IEEE) plays a leading role in developing high-quality standards for computational sensing technologies through the working groups of the Synthetic Aperture Standards Committee (SASC).  In this column we highlight the standards activities of the P3383 Performance Metrics for Integrated Sensing and Communication (ISAC) Systems Working Group and the P3343 Spatio-Temporal Synchronization of a Synthetic Aperture of Distributed Sensors Working Group.    
\end{abstract}

%
%
%
%
%
\section{Computational Sensing and the Importance of Standards}
\label{sec:intro}
Distributed sensors, such as radar or sonar, are often networked together to improve detection performance.  In a networked configuration, these sensors may share detection reports over a wireless link and coordinate target tracks so as to leverage the benefits of spatial diversity and different aspect angles to an object.  Even greater performance benefits can be accrued if the distributed sensors are synchronized.  In this case, the sensors can function collectively as a single aperture and perform tasks that require a high degree of spatial and temporal coherence such as imaging.

Sensing functionality can also be implemented in a traditional communication system by incorporating waveforms that exhibit low autocorrelation sidelobes or by time sharing the transmit/receive aperture between different tasks.  If a channel sounding capability is available as well, then it may facilitate sensing performance in a communication system by measuring the scattering due to objects in the scene.  The additional degrees of freedom leveraged by optimizing waveform design, managing aperture resources, or by making channel measurements can form the basis for simultaneously transmitting data while sensing objects.

Integrated sensing and communications (ISAC) and the spatio-temporal synchronization of distributed apertures are two key technologies within the scope of the IEEE Signal Processing Society's Synthetic Aperture Standards Committee (SPS-SASC).  More specifically, the P3383 Performance Metrics for Integrated Sensing and Communication (ISAC) Systems Working Group and the P3343 Spatio-Temporal Synchronization of a Synthetic Aperture of Distributed Sensors Working Group are developing standards for these technologies.

Standards are technical documents that define industry best practices and allow systems manufactured by different vendors to be interoperable.  One essential difference between a standard and an academic journal paper is that the process used to develop a standard is transparent and auditable.  In other words, standards developing organizations (SDOs), such as the IEEE, strive to create standards through a consensus process with input from a broad community.  Differing viewpoints and perspectives are discussed during the authorship process such that the best technology solutions emerge from a working group's deliberations.  Strict protocols adhered to during the meetings of a working group avoid dominance by a single viewpoint or conflicts of interest by the voting members.  These precautions ensure that the requirements stipulated in a published standard are trustworthy and well-validated.

Modern commerce and government acquisitions rely extensively on standards for defining product quality and vendor performance.  Standards are also essential in the system engineering community for creating digital twins since the attributes of a simulated system may be derived entirely from standards.  This column is the second part in a series that highlights the standards activities of the SASC and in this installment we focus on the P3383 and P3343 working groups that are developing standards for ISAC and the spatio-temporal synchronization of distributed apertures.

%

%
\section{Performance Metrics for Integrated Sensing and Communication (ISAC) Systems}
\label{sec:ISAC}

ISAC refers to approaches that combine radar and communication functions. 
Although the sensing part is an estimation and/or detection process, the actual implementation varies considerably among the proposed methods. 
This renders a fair comparison solely based on common metrics such as bit error rate (BER) or Cramer Rao lower bound (CRLB) hard to achieve.
Furthermore, another important aspect of ISAC is also signal/waveform design, where signal processing and information theoretical aspects enter the fray. 
Both the estimation and signal design perspectives need to be given suitable metrics as a foundation not only to compare the methods but also to drive the method development into suitable directions. 

This section looks into possible approaches for defining  such metrics starting from the discussion of methods for estimation and the design of signals. 
While there exists a distinction between communication-centric and radar-centric ISAC techniques, i.e., whether standard communication waveforms are utilized to also achieve a sensing functionality, or vice versa, the discussion based on this distinction is omitted and an agnostic perspective is taken since, in the end, it is the choice of the user where the focus is placed. 

The structure of this section is as follows.
First, ISAC is contextualized within the broader scope of sensing and a brief historical outline is given. 
Then, a focused summary of the various methods and signal design techniques is given. 
Based thereon, an in-depth motivation for new or advanced metrics are given. 
Finally, candidate criteria for such metrics are discussed, subdivided by criteria for signal design and estimation methods for clarity of exposition.   

\subsection{Brief Description of Strategies in ISAC Architectures}
\label{subsec:ISAC_localization}

With the goal of facilitating a parallel operation of both radar and communications functionalities, ISAC is similar to co-existence, passive sensing (passive radar), and cognitive radio but differs in the sense that, radar and communications are integrated into one joint system -- most importantly with a fundamental single ``ISAC" waveform, see, e.g., \cite{Wu.2023} for a comparison between these system types. 
Both the radar and communications communities independently embarked on this idea for ISAC. 

On the communications side, the OFDM-based approach proposed by \cite{Sturm.2011} in 2011 is often mentioned as a starting point, leveraging this standard waveform of communications to extract the inherent radar information of the scatterers of the environment. 
In this context, the application of signals based on the IEEE 802.11 standard is prominently mentioned; see \cite{Wu.2023,Du.2025} and references therein. 
Based thereon, a variety of next-generation waveforms are investigated, which better encapsulate the double-dispersivity of the wireless channel (time- and frequency-selectivity) \cite{Rou_SPM_2024}. 
Among these waveforms are the recently popularized Orthogonal Time Frequency Space (OTFS) \cite{Hadani_WCNC17}, Affine Frequency Division Mutliplexing (AFDM) waveforms \cite{Bemani_TWC23}, and others \cite{lin2022orthogonal,senger2025affine} from which methods enabling ISAC functionalities have been thoroughly investigated \cite{Ranasinghe_TWC_2025}.
Of course, there also exist a plethora of approaches leveraging the inherent information of the channels to extract environmental information \cite{Liu_ISAC1,Cui_ISAC2,Zhang_ISAC3,Rou_TWC24,Rou_Asilomar22}.

On the radar side, the interest in waveforms that also incorporate communication symbols started in the same time frame, see, e.g., \cite{Blunt.2010,Hassanien.2018,Gaglione.2018,Mai_JRC23} and  \cite{deOliveira.2022} for an overview. 
A first idea for using radar to also transmit data can already be found  in 1978 with \cite{Cager.1978}. 
These two starting points gave rise to a myriad of names, like RadCom, Dual-Function Radar Communications, Joint Communications and Sensing, or Joint Radar (and) Communications, essentially referring to the same idea of ISAC but emphasizing that the focus is either on the radar or communications.  

In recent years, the interest in ISAC has soared and is currently mainly driven by the communications side with the development of the sixth-generation (6G) of wireless communications systems \cite{Wild.2021}. 
With respect to the development as a function within 6G, it should be mentioned that 5G and earlier generations already included positioning capabilities, i.e., signals and processes for the localization of user equipment, see \cite{Italiano.2024} for a detailed discussion. 
Hence, ISAC should be not confused with positioning as the radar-like measurement incorporated in ISAC provides the capability to perform localization of uncooperative objects, i.e., objects that are not linked to the network. 
In \cite{Liu.2022}, this distinction is made through the terms \textit{device-based} and \textit{device-free}. 

Similar to the distinction with respect to classical positioning, a differentiation of ISAC with respect to (classical) radar can also be made. 
While the underlying principle of the sensing functionality of ISAC and radar are the same, the implementation of this sensing capability is different. 
Both estimate the parameters of objects in the scene based on the same physical phenomena, i.e., the path delay and Doppler shift of the propagated electromagnetic wave due to target range and velocity, and the angle-of-arrival due to target bearing.
But, the term \textit{radar} is usually linked to an operation based on (in terms of their modulation) simple, high-bandwidth signals and a specific type of estimation -- matched filtering.
Sensing under an ISAC paradigm, however, evolves into a more general estimation problem, different from matched filtering, due to the constraints imposed by the communication functionality of the signal, which in turn gives rise to more complex estimation schemes taking partially the transmitted data into account. 
Also, the communications part of ISAC creates an inherently networked (multi-agent) system, which can not only be leveraged for multi-static sensing in terms of transmit and receive but also for a distributed computation.
Especially, new ISAC waveforms, which are designed to satisfy high communication performance criteria, are not optimally designed for matched filtering-based radar estimation since their ambiguity function will differ from the desired spike-like form without sidelobes. 

This point of view of considering ISAC in terms of an estimation problem (and not a radar-like problem) is further strengthened by \cite{Liu.2023}, where the \textit{signals-and-systems duality} of sensing as well as communications is pointed out.
In both cases, the received signal is generated through a convolution of a known signal (sensing: transmitted signal, communications: channel) with an unknown signal (sensing: channel, communications: transmitted signal) and the estimation of the unknown signal. 
Hence, sensing can, as shown in \cite{Bell.1993,Chiriyath.2017}, also be treated from an information theoretical point of view, implicitly linking both sensing and communications into a single joint problem.

Although the field experienced a rapid growth in the last ten years, there are still open research objectives, like synchronization or full-duplex transceivers, see, e.g., the overview in \cite[Chapter 1]{Wu.2023}. 
One crucial aspect is a metric to achieve comparability between different estimation approaches and signals. 
Estimators and signals should be considered separately since 
methods are designed to extract the information of the ``system'', i.e., channel, from a signal, while signals are designed to transport and represent this information. 
Therefore, while both tasks jointly serve the purpose of parameter estimation, the focus is different.

\subsection{Outline of Approaches to the Problem of Integrated Sensing and Communication} \label{sec:ISAC:OutlineISAC}

\subsubsection{Estimators \& Signals}\label{sec:ISAC:OutlineISAC:MethodsSignals}

Focusing on the estimators for ISAC, one of the most common fundamental techniques is the Maximum Likelihood Estimator (MLE) \cite{Zhang.2021} which provides asymptotically optimal performance with a high computational complexity.
Another popular method is the use of Compressed Sensing (CS) or Least Squares (LS) \cite{Liu.2020} to extract parameters, with the former being superior in settings with sparsity.
For example, \cite{RanasingheARXIV_blind_2025} proposes a blind bistatic radar parameter estimation technique, which enables ISAC by allowing passive (receive) base
stations (BSs) to extract sensing parameters (ranges and velocities of targets), without requiring knowledge of the information sent by an active (transmit) BS to its users.
This is done via a covariance-based formulation of the doubly-dispersive environment which is recast and solved via the well-known fractional programming technique.
The authors in \cite{Ranasinghe_TWC_2025} propose two belief propagation-based algorithms to solve the ISAC problem by transforming the sensing aspect into a sparse CS-based problem.
Considering the underlying concept of Compressed Sensing, the aspect of random sampling of the sensing matrix (representing the signal model or dictionary) as described in \cite{Foucart.2013} allows for gaps, i.e., arbitrary sampling in the respective domain (frequency for the delay, time for Doppler), and thus gaps in the spectrum. 
Finally, subspace-based methods leveraging techniques such as MUSIC and ESPRIT \cite{Su.2025} are also popular for ISAC, specially in the cases where the AoA's and AoDs have to estimated with a high resolution \cite{Ranasinghe_ICASSP_2024}.

Dual-function waveform design in ISAC can broadly be categorized into communication-centric and radar-centric approaches, each prioritizing different design goals.
Communication-centric waveforms, such as Orthogonal Time Frequency Space (OTFS) and Affine Frequency Division Multiplexing (AFDM), originally intended to handle doubly-dispersive channels for robust communication, have recently been adapted for radar sensing due to their advantageous ambiguity properties \cite{Rou_SPM_2024}. 
Specifically, OTFS waveforms \cite{Hadani_WCNC17}, structured in the delay-Doppler domain, exhibit robustness against Doppler shifts and multipath fading, thus making them suitable for high-performance radar sensing with relatively straightforward modifications. Indeed, Gaudio \textit{et al.} \cite{Gaudio2019} demonstrated that OTFS-based ISAC systems achieve a Cramér-Rao lower bound (CRLB) comparable to frequency-modulated continuous-wave (FMCW) radars, underscoring their suitability for joint applications.

On the radar-centric side, waveforms traditionally designed for radar, such as chirps and phase-coded signals, embed communication data within the radar pulses or modulation schemes \cite{Liu_JSAC_2022-1,Liu_JSAC_2022-2}. Liu \textit{et al.} \cite{Liu_JSAC_2022-1,Liu_JSAC_2022-2} extensively analyzed radar-centric ISAC waveforms, proposing advanced modulation strategies that maintain radar detection capabilities while providing reliable communication performance through subtle waveform embedding. 
These radar-centric waveforms typically involve modifications to traditional radar signals, embedding symbols by adjusting phase, amplitude, or frequency modulations while aiming to preserve essential radar features like low ambiguity sidelobes and accurate Doppler resolution.

Fundamentally, the dual-function waveform design reveals the intrinsic trade-offs between radar sensing and communication objectives—communication signals prioritize spectral efficiency, symbol recovery, and error resilience, while radar waveforms emphasize ambiguity function optimization, detection sensitivity, and parameter estimation accuracy. 
Thus, effective ISAC waveform design must carefully balance these divergent requirements, which necessitates joint optimization frameworks and novel metric definitions that capture the intricacies of both sensing and communications.

  \subsubsection{Commonly used metrics}
  \label{sec:ISAC:OutlineISAC:Mainmetrics}
    Similar to the classification into estimators and signal, the commonly used (theoretical) metrics of Cramér-Rao Lower Bound (CRLB), Mean Square Error (MSE), Bit Error Rate (BER) or Symbol Error Rate (SER), Channel Capacity and Mutual Information (MI) can also be subdivided into estimator metrics and signal metrics. As for the estimators of the sensing parameters $\theta$ defining the linear time-variant impulse response 
    \begin{equation}
      g(t,\tau) = \sum_{p=0}^{P-1}h_p\delta(\tau-\tau_p)e^{j2\pi f \tau_p} 
    \end{equation}
    of the doubly-dispersive (or delay Doppler) channel\footnote{This doubly-dispersive representation of the channel is the system theoretical description of the two dimensional fast time/slow time treatment of the data in radar. Also, note that a Doppler delay channel is different from a delay Doppler channel in terms of the position where the Doppler takes effect (at the input or the output), see \cite{Bello.1963}.}, it is MSE and the CRLB. In terms of the ensemble mean, the MSE is defined by 
    \begin{equation}
      \sigma^2_\mathrm{MSE} = E\left((\theta - \hat{\theta}_y)(\theta - \hat{\theta}_y)^\mathrm{H}\right) \; ,
    \end{equation}
    where the (stochastic) variable $\theta$ represents the true value, the (stochastic) variable $\hat{\theta}(y)$ represents the estimate based on measured values $y$, and $E(\cdot)$ is the expected value. In terms of the sample mean, it is defined by
    \begin{equation*}
      \hat{\sigma}^2_\mathrm{MSE} = \frac{1}{N}\sum_{k=0}^{N-1}(\theta - \hat{\theta}_y(k))(\theta(k) - \hat{\theta}_y(k))^\mathrm{H} \; ,
    \end{equation*}
    for $N$ estimations with varying datasets $y$. The CRLB is defined by
    \begin{equation}
      \sigma^2_\mathrm{MSE} \geq \sigma^2_{CRLB} = {E\left(\left(\frac{\partial p_y(\theta|y)}{\partial\theta}\right)^2\right)}^{-1}
      = {-E\left(\frac{\partial^2 p_y(\theta|y)}{\partial\theta^2}\right)}^{-1}
    \end{equation}
    where $p_y(\theta|y)$ is the probability density function of $\theta$ given $y$ or, vice versa, the likelihood function of $y$ given $\theta$. The term $-E\left(\frac{\partial^2 p_y(y|\theta)}{\partial\theta^2}\right)$ is called the Fisher Information Matrix. 
  
    For communications, the most popular metric is the BER defined by
    \begin{equation}
      \text{BER} = \frac{b_\text{errors}}{b},
    \end{equation}
    where $b_\text{errors}$ denotes the number of bit errors observed and $b$ is the total number of bits transmitted.
    The symbol error rate (SER), which counts the total number of symbol errors instead of the bit errors, is then related to the BER through the number of bits per symbol.
    Theoretical BERs can also be derived for specific constellations under additive white Gaussian noise (AWGN) channels of which a very common example is Binary Phase Shift Keying (BPSK), whose BER is given as
    \begin{equation}
      \label{eq:BER_BPSK}
      \text{BER}_\text{BPSK} = \text{SER}_\text{BPSK} = \mathcal{Q} \bigg( \sqrt{\frac{2E_b}{P_{nn}}} \bigg) \; ,
    \end{equation}
    where $\mathcal{Q}(.)$ denotes the Gaussian Q-function, $E_b$ is the energy per bit and $P_{nn}$ is the noise power spectral density (assumed to be constant over frequency). Notice that for BPSK, the BER is identical to the SER.
  
    The second commonly used metric is the channel capacity between the input $u$ and output $y$ of a channel defined by
    \begin{equation}\label{eq:ChannelCapacity}
       C = \max_{p_{u}}I(u,y)\; , 
    \end{equation}
    where $I$ is the Mutual Information given (for discrete probability density functions) by
    \begin{equation}\label{eq:MIComm}
       I(X,Y) = \sum_{x\in\mathcal{X}}\sum_{y \in \mathcal{Y}}p_{X,Y}(x,y)\log\left(\frac{p_{X,Y}(x,y)}{p_{X}(x)p_{Y}(y)}\right) \; .
    \end{equation}  
    The communication criteria are anchored in the detection problem, which however is also relevant as a secondary step for the sensing, i.e., the detection of targets among the estimated parameters. For a more in-depth discussion of the criteria see \cite{Richards.2014,Liu.2022}, where it is also shown that the bandwidth of a signal and its temporal length are also limiting factors for the CRLB of the range/delay and Doppler estimation.
  
    As discussed in \cite{Bell.1993}, sensing can also be set in to the context of information theory and be based on the MI. In this case, the mutual information between the impulse response of the channel $g(t)$, i.e., the target parameters, and the output signal $y(t)$ conditional to the input $u(t)$, i.e., $I(g(t),y(t)|u(t))$ is considered. For an undisturbed output $z(t) = (g \ast u)(t)$ of the system with $y(t) = z(t) + n(t)$, $z(t)$ is considered to be equivalent to $g(t)$ and hence this conditional MI is given by
    \begin{equation}\label{eq:conditionalMI}
      I(z(t),y(t)|u(t)) = T\int_\mathcal{W}  \ln\left(1 + \frac{2|u(f)|^2\sigma_g^2(f)}{P_{nn}(f)T}\right)df \; ,
    \end{equation}
    where $T$ is the time interval, which the signals are considered over, $\mathcal{W}$ is the frequency interval, which the signals are confined to, $\sigma_g^2(f)$ is the spectral variance of the impulse response, which is here assumed to be random due to a Gaussian target distribution, and $P_{nn}(f)$ is the spectral power density of the noise. This defines a design criterion for $u(t)$. As discussed in \cite{Bell.1993}, $I(z(t),y(t)|u(t))$ is to be maximized based on $u(t)$. Another signal criterion, which is used for the design of radar waveforms, is the ambiguity function. It is essentially the cross-correlation function of the input signal with a delayed and Doppler modulated version of itself. For a scalar signal, it is defined as, e.g.,\cite{Richards.2014} 
    \begin{equation}\label{eq:Ambiguity}
      A(\nu,\tau) = \left|\int_{-\infty}^{\infty}u(t)e^{j2\pi\nu t}u^\ast(t-\tau)dt\right|
                = \left|\int_{-\infty}^{\infty}u^\ast(f)u(f-\nu)e^{j2\pi ft}df\right| \;,
    \end{equation}
    where $\tau$ defines a delay and $\nu$ the Doppler. It has its maximum at $A(0,0)$, which is equal to the energy of the signal, while the total energy under the surface is equal to $(A(0,0))^2$. In the context of the ambiguity function, it should be noted that it is directly linked to the correlation approach of a matched filter. That is, the ambiguity function as a metric for the quality of the signal loses its informative value, the further away an estimator is from a correlation approach. In CS, for example, the information represented by the ambiguity function reappears in terms of the Gramian of a sensing matrix, from which the coherence of the sensing matrix is determined. While this similarity measure is important, it is only a part of the overall estimation and not the pivotal element.
  
    Hence, while there are useful and elaborate metrics for evaluating the estimators and signals, they are either based on a-priori knowledge, which is hard to come-by in a practical scenario, e.g., the probability density function, or too simple to reflect the whole estimation process as they only concentrate on one aspect, e.g., the MSE or CRLB provide information on the estimation error, but not the complexity of the estimator. This issue requires the development of a suitable, practical usable metric, which incorporates multiple aspects of the estimation process. For the signal design, the situation is simpler as it is reasonable to design a signal based on certain assumptions regarding the noise and the system.
  
  \subsection{Motivation for a general metric} \label{sec:ISAC:motiv}
    Due to the variety of the methods, a metric for a meaningful statement with respect of the quality of the methods needs to take multiple aspects  into account. Thus, a unifying metric is needed. The underlying idea and motivation for such an approach is outlined in this sub-section.

    \subsubsection{Different methods, different signals, same goal}
      In Section \ref{sec:ISAC:OutlineISAC:MethodsSignals}, the different methods were outlined. 
      While all have eventually the same goal of the estimation (and detection) of the target signal\footnote{This time-frequency representation of the delay-Doppler channel (time over Doppler domain, frequency over delay domain) is just used to emphasize the underlying estimation of complex exponentials, see \cite{Bello.1963} for an in-depth discussion of the different representations of delay-Doppler channel.}
      \begin{equation}\label{eq:DDchannelmodel}
        g(t,f) = \sum_{p=0}^{P-1}h_pe^{j2\pi t \nu_p}e^{j2\pi f \tau_p} 
      \end{equation}
      for the delays $\tau_p$, Dopplers $\nu_p$, and complex amplitudes ${h_p}$, the underlying structures differ. 
      For example, while Fourier-based approaches, which require an equidistant sampling in time, calculate $g(t,f)$ not in terms of the above defining components, i.e., a parametric model, but rather in terms of the non-parametric impulse response. 
      Algorithms like MUSIC and ESPRIT are designed to extract directly the complex exponentials from a noisy signal $g(t,f) + n(t,f)$ where the frequencies (here the delays $\tau_p$ and Dopplers $\nu_p$) are allowed to be continuous. 
      Yet again, Compressed Sensing defines an estimation problem based on a (discrete) dictionary\footnote{Approaches for continuous dictionaries or dictionary-free approaches are also designed since discrete dictionaries create the problem of basis mismatch, e.g., \cite{Austin.2013,Tang.2013}.} of possible candidate signals that might establish $g(t,f)$.
      That is, opposed to the Fourier-based approach, the latter two classes of algorithms assume (parametric) models for $g(t,f)$ in the estimation. 
      These estimations based on models include one further aspect -- they also perform the detection step, i.e., the decision on which signal component is part of the signal. 
      This is especially evident for Compressed Sensing, where the (assumed) components of $g(t,f)$ are picked from the dictionary. 
      Also, the possibility of discontinuous spectra, i.e., gaps over $t$ or $f$, makes the estimation over larges bandwidths (in frequency and time)  possible, while the actually occupied resources remain limited, see e.g. \cite{Bathelt.2023}. 
      Yet, the advantages over Fourier-based approaches come with higher computational complexity and/or the need to have an a-priori knowledge or an estimate of the number of targets $P$. 

      Theses points showcase the underlying problems in terms of comparability of the different approaches and many more can be found.
      Furthermore, the aspect of communication-centric and radar-centric approaches to the problem add another layer since the signals (especially in terms of their bandwidth or occupied spectrum) differ for both approaches. 
      These aspects can be generalized with regard to the available resources, i.e., spectrum and computation.
   
    \subsubsection{Is the estimation error to be taken into account?}
      With the short discussion above, the answer to this question is obvious. 
      If a fair comparison of the different estimation approaches is sought after, then the estimation error is not the only criterion which is needed. 
      Also the cost in terms of spectrum, computational load, additional calculations, etc. needs to be taken into account. 
      For a genuine practical scenario the ground truth of the measured scene is hard to come by. 
      In such a case, only the recorded data is known. 
      In the case of a simulation study, the ground truth is known and can be used. 
      For communication-centric ISAC there is an additional limitation imposed by the signal structure due to the standards, e.g., 3GPP, i.e, how the frames, resource blocks, and reference symbols therein are distributed. 

      Thus, a metric for the estimator needs to take the available resources or costs into account and not only focus on the result.
      It should furthermore not be based on theoretically derived metrics like CRLB since there is rarely usable a-priori knowledge on the noise and the AWGN assumption (which is often used to derive the CRLB) is quite limiting, since the noise is usually correlated. 
      If a metric for the signal design is considered, it is however conceivable to work based on the (conditional) MI. 
      For the signal design some assumptions need to be made. 
      But here again, a more realistic point of view is to be taken when the disturbance is considered, i.e., there is not only white noise but also clutter from unwanted targets, which can be assumed to be statistically different from the desired targets.

  \subsection{Ideas from System Identification}
  \label{subsec:IdeasSysId}
    In Section \ref{subsec:ISAC_localization}, the perspective of the channel from the point of view of a ``system'' as discussed by \cite{Liu.2023} was briefly mentioned and, in Section \ref{sec:ISAC:OutlineISAC:MethodsSignals}, the perspective of parametric and non-parametric models was taken. 
    Hence, it is possible to draw comparisons between ISAC and system identification.
    The similarities are with respect to the modeling of a system either in a parametric or non-parametric way -- for ISAC it is structurally a two-dimensional moving average filter, compare with the discussion on model structure in \cite{Ljung.2009}. 
    But, the primary focus of system identification is on black-box models for description of the input-output relation of dynamic systems where the real-valued coefficients of a set of different model structures for difference or differential equations are to be estimated. 
    In the case of ISAC, it is a a physical model, where not only the coefficients but also the delay and Doppler parameters\footnote{For arrays, also angle of arrival and angle of departure.} are sought-after.
    That is, in fact, the model structure itself is to be identified.

    Nevertheless, certain ideas and concepts can be transferred.
    For the evaluation of the model quality, two metrics can be considered here -- the coefficient of determination $R^2$ and the Akaike's Final Prediction Error, see \cite{Kroll.2014,Ljung.2009}.
    The coefficient of determination is given by
    \begin{equation}
        R^2 = 1 - \frac{\sum_{k=0}^{N-1}\left(y(k)-\hat{y}(k)\right)^2}{\sum_{k=0}^{N-1}\left(y(k)-\overline{y}(k)\right)^2} \; ,
    \end{equation}
    where $N$ is the number of recorded data points and $\overline{y}$ is the mean of $y$.
    It evaluates how much of the variance of the output of a system is represented by the output (prediction) of the model ${\hat{y}}$ and is thus also called \textit{variance accounted for}.
    That is, it compares the models solely based on the quality of representing the output, where the best score is $1$ (for negative scores the value is set to zero).
    The Final Prediction Error criterion is given by
    \begin{equation}\label{eq:FPE}
      J_{FPE} = \frac{1+\frac{d_\mathcal{M}}{N}}{1-\frac{d_\mathcal{M}}{N}}\frac{1}{N}\sum_{k=0}^{N-1}\left(y(k)-\hat{y}(k)\right)^2 \; ,
    \end{equation}
    where $d_M$ is the model dimension, and, as descried by \cite{Ljung.2009}, takes the \textit{cost of (the number of) the parameters} into account. 
    A more general expression for this is given by \cite[(16.42)]{Ljung.2009}
    \begin{equation}\label{eq:CostCriterion}
       W_N(\theta,\mathcal{M}) = (1+U_N(\mathcal{M}))\frac{1}{N}\sum_{k=0}^{N-1}\ell(y(k)-\hat{y}(k)) \; , 
    \end{equation}
    where $\ell$ is the norm used for the prediction error $y(k)-\hat{y}(k)$ and $U_N(\mathcal{M})$ is a penalty term for the model structure.
    While the number of parameters is fixed in the ISAC case due to the underlying physical meaning of the (number of) parameters, the idea of penalizing the estimation error by the cost of achieving this error is noteworthy for an ISAC metric.

    Finally, signal design is also a relevant question for system identification.
    Here, the concept of \textit{informative experiments} is discussed.
    The goal is to define a signal such that all properties of the system expose itself in the output signal and to facilitate a distinction between any two models of the system solely based on the characteristic of the estimation error.
    That is, each possible model has to have its own unique sequence of estimation errors.
    This leads to the structural signal criterion of persistence of excitation, which is given in terms of
    \begin{equation}
      0 = |\Delta G(f)|^2\Phi_{uu}(f) \; ,
    \end{equation}
    where $\Delta G(f)$ is the difference between any two models and $\Phi_{uu}(f)$ is the power spectral density of the input. 
    A signal is said to informative, if this equation holds only if $\Delta G(f) = 0$.
    Vice versa, it is not informative, if the input $u$ can be filtered to zero by $\Delta G(f)$.
    This leads to a structural criterion in terms of how large the frequency support needs to be for $u$.
    Following the underlying idea a similar result might be given for $g(t,f)$.
    
  \subsection{A set of criteria for ISAC metrics} \label{sec:ISAC:Criteria}
    When considering metrics for ISAC, the general questions are with respect to
    \begin{itemize}
        \item Which components of the estimation process need a metric?
        \item Which aspects of these components need the metric to be based on?
    \end{itemize}
    The former question is easily answered. From the discussion in Section \ref{sec:ISAC:OutlineISAC}, the rough division should be between signals and estimation approaches. The latter question depends on reasonable assumptions and knowledge that are available for the process.

    \subsubsection{Signal criteria}
      With respect to the signal design, it is reasonable to assume that the signal is designed for a certain scenario. Hence, the conditional MI \eqref{eq:conditionalMI} presents an already well-established basic criterion for signals in terms of the sensing capabilities. On the communication side, the channel capacity \eqref{eq:ChannelCapacity} or rather mutual information \eqref{eq:MIComm} establish a criterion for the signal in terms of its communication capabilities. Since the joint distribution $p_{X,Y}(x,y)$ is also given by $p_{Y|X}(y|x)p_X(x)$, where $p_{Y|X}(y|x)$ is essentially a model for the channel, \eqref{eq:MIComm} provides also a measure for communication capabilities given a certain scenario. The joint metric should also take the need of the application into account, i.e., whether the main focus is on sensing or communication. Hence, such a signal metric can be given the basic structure
      \begin{equation}\label{eq:ISAC_signal_premetric}
        J(u) =  \lambda I_s(z,y|u) + (1-\lambda)I_c(u,y(u)) \; ,
      \end{equation}
      where $\lambda \in (0,1)$ serves as a user-chosen weighting between sensing $I_s$ and communications $I_c$. As for $I_c$, the notation $y(u)$ reflects the fact that for an assumed scenario the output signal $y$ can be determined based on the input signal. For \eqref{eq:ISAC_signal_premetric} to become fully usable, some further considerations and derivations are needed, e.g., both $I_s$ and $I_c$ should have the same magnitude to avoid one of them dominating the metric. Furthermore, other aspects could be included:
      \begin{itemize}
        \item \textit{Structural capacity for sensing:} Beyond \eqref{eq:conditionalMI}, the idea of informative experiments as outlined in Section \ref{subsec:IdeasSysId} should be included to balance the sensing/communication capabilities against the minimal spectral costs for achieving theses capabilities. 
        \item \textit{Robustness against clutter:} Dealing with clutter, i.e., unwanted signal returns, is also important under an ISAC paradigm. Hence, the spectral power density of the noise/disturbance $P_{nn}$ of $\eqref{eq:conditionalMI}$ should give a suitable representation of the clutter environment such that a signal is also measured against its capabilities to extract the wanted signal returns within clutter. Another approach would be to introduce a penalty term similar to $I_s$ focusing on the return due to clutter. The smaller the return from clutter is from the outset, the less it has to be dealt with during the estimation/detection stages.
        \item \textit{Signal ambiguity:} Depending on the intended estimation the ambiguity (frequency-modulated auto-correlation) of the signal might introduce issues within the estimation/detection stages. Hence, the side lobe level (relative to the main lobe) as revealed by the ambiguity function \eqref{eq:Ambiguity} might be taken into consideration.
        \item \textit{Peak-to-Average-Power Ratio:} The better the Peak-to-Average-Power Ratio or Crest Factor of a signal is the higher the signal power can be without driving amplifiers into saturation. This in turn is relevant for the maximum range, which the sensing system can achieve with the signal, since the maximum range is (among other factors) proportional to the transmitted power.
      \end{itemize}

    \subsubsection{Estimator criteria}
      With respect to the estimators, not many assumptions can be made in a practical scenario -- in a real world scenario, the ground truth might not be known or the model of point scatterers is too idealized. If simulation studies or studies in a lab environment are considered, the true target parameters are however known. Using the ideas from Section \ref{subsec:IdeasSysId} a heuristically basic structure for an estimator metric can be given by
      \begin{equation}\label{eq:EstimatorProtoMetric}
        J = w_\text{cost}\left(\lambda\frac{1}{K}\sum_{k=0}^{K-1}\left(\phi(k)-\hat{\phi}(k)\right)^2 + (1-\lambda)\frac{b_\text{errors}}{b} \right) \; ,
      \end{equation}
      where $\lambda$ is again a balancing weight between the importance of the sensing error (first term) and the communication error (second term, here as BER), $\phi$ and $\hat{\phi}$ are the compared entities for sensing, i.e., the true and estimated parameters $\theta$ and $\hat{\theta}$ or the measured and predicted data $y$ and $\hat{y}$ (predicted by the estimated model, here scalar), $K$ is hence either the number of parameters $P$ or the number of data points $N$, and $w_\text{cost}$ is a weighting due to the cost of the individual estimator. This cost weighting can be structured similar to \eqref{eq:FPE} or \eqref{eq:CostCriterion}. However, for ISAC and the different estimation approaches, the cost is not defined by the number of parameters but the employed resources to achieve the respective sensing (and communication) errors. Thus, instead of $d_\mathcal{M}$ the used resources have to be used so that $w_\text{cost}$ similar to \eqref{eq:FPE} or \eqref{eq:CostCriterion} can be defined as
      \begin{equation}
         w_\text{cost} = \frac{1+\frac{1}{C_\text{max}}\sum_{k=1}^{K}\lambda_{c,k}C_k}{1-\frac{1}{C_\text{max}}\sum_{k=1}^{K}\lambda_{c,k}C_k}  \qquad w_\text{cost} = 1 + \sum_{k=1}^{K}\lambda_{c,k}C_k \; ,
      \end{equation}
      where $\lambda_{c,k}$ with $\sum_{k=1}^K\lambda_{c,k} = 1$ are user chosen weights for the importance of the individual cost element $C_k$ and $C_\text{max}$ is the maximal cost so that for the FPE-like weight $w_\text{cost} \geq 1$. Essential and directly numerically quantifiable aspects are  
      \begin{itemize}
        \item \textit{Data and spectrum usage} Number of sample points in time $N$ and occupied spectrum (e.g., in terms of spectral sampling points, number of sub-carrier, or bandwidth) used for achieving a certain estimation quality reflects ``How well does a method extract information on the channel from the data?''.
        \item \textit{Computational load} For the computational load, the big O notation is the first rough estimate. Since usually only the core operations that scale with the number of inputs are considered and the big O notation evaluates the asymptotic behavior, a more detailed metric for the computational load might be used which also catches the sub routines of the estimators. 
      \end{itemize}        
      Other aspects, which might be considered but are not directly numerically quantifiable, are (for the above metric or in general for an evaluation of methods)
      \begin{itemize}
        \item \textit{Resolution} Different methods may provide different levels of attainable resolution for the parameters. Precision and accuracy map to the estimation error (if $\phi$ equals $\theta$) or prediction error (if $\phi$ equals $y$).
        \item \textit{Scalability} This is related to the aspect of the quantity (unknown number) of targets. Certain approaches need a-priori information on the number of targets, which might not be a problem for proof-of-concept experiments, but can become an estimation problem of its own when the number of targets or the complexity of the scenario (e.g., clutter or shape of the targets) increase. This can be also extended to the question whether the methods can operate properly when the assumed disturbance is different (AWGN or clutter).
        \item \textit{Model-based approach or model-free approach} Depending on the approaches, method might be based on an assumed model or not, e.g., a matched filter effectively calculates the impulse response and Compressed Sensing is based on a signal model. In the context of model-based approaches, such a method combines the estimation with the detection of the respective signal components. The detection is not a subsequent step as in the case of a matched filter approach. Such approaches are however prone to model mismatch. For example, the Doppler signature of a target in terms of micro-Doppler might not be reflected as desired if the model, e.g, \eqref{eq:DDchannelmodel}, does not capture this aspect because an insufficient Doppler grid resolution was chosen. 
        \item \textit{Robustness} This aspect is linked to the previous two aspects and focuses on the robustness, i.e., impact on the error in the case of deviation from the assumptions of the model, e.g., signal model, noise model.   
        \item \textit{Multi-static or mono-static} Sensing can be done in a multi-static or mono-static set-up. Similar to scalability, it is important to asses whether a method can be used in both set-ups or is specifically designed for one of the two set-ups. Mono-static sensing needs some intrinsic self-interference cancellation if full-duplex is unavailable in communication-centric ISAC, whereas multi-static sensing requires precise (time) synchronization. Both are additional costs for the estimation.
        \item \textit{Coherence} Processing the signals in-phase, i.e., coherently, can improve the probability of detection. Hence, if the data is available only in a non-coherent way (i.e., if only envelope detection is possible), the susceptibility to errors of the results might be assessed. 
      \end{itemize}
      With respect to the usage of the data $y$ in \eqref{eq:EstimatorProtoMetric}, it should be noted that, in the case of a significant clutter component and an estimation which focuses only on the targets of interest, the remaining clutter component will dominate the error $\frac{1}{N}\sum_{k=0}^{N-1}\left(y(k)-\hat{y}(k)\right)^2$. Thus, it might be conveived to estimate the clutter as well (similar to a disturbance model), and include its estimation error as well, for example when validation data sets are used, so that $\frac{1}{N}\sum_{k=0}^{N-1}\left(y(k)-\hat{y}(k)\right)^2$ is again only an estimation error.


%

%
\section{Spatiotemporal Synchronization of Distributed Apertures}
\label{sec:spatiotemporal-sync}
Distributed MIMO (D-MIMO), also known as cell-free MIMO, is a technology that is gaining traction 
as a key radio architecture for future generations of mobile communications standards~\cite{Ngo24ultradenseDMIMO}.
The inherent macro-diversity and channel hardening effect in distributed radio infrastructures make D-MIMO an enabling technology for ultra-reliable low-latency communication (URLLC)~\cite[Sec.\,VI\,D)]{Nelson25DMIMO}.
When reciprocity-based beamforming is applied in the uplink of a time-division-duplex (TDD) system, only coarse synchronization---i.e., aligning symbol timings within the length of the OFDM cyclic prefix~\cite[Sec.\,V.\,F]{Ngo24ultradenseDMIMO} and bounding the carrier-frequency offset---is required, because both impairments can be estimated and removed in baseband.
In the context of wireless power transfer (WPT), enlarging and distributing the antenna infrastructure further concentrates power on the intended receiving node, aiding radiation safety and regulatory compliance~\cite{Deutschmann25WCM}.
Downlink beamforming requires tight phase calibration~\cite{Larsson23PhaseCalibration} of each node's RF chain before coherent joint transmission (CJT)~\cite{Omer23CJT}. 
With the emergence of D-MIMO, multistatic ISAC---where spatially separated transmit–receive pairs jointly deliver data and perform radar sensing---has likewise become a popular research frontier with significant performance and robustness gains inherently available in the distributed nature of the infrastructure~\cite{Guo24DMIMO_ISAC,Deutschmann24SPAWC}. 
In addition to knowledge of the \textit{temporal states}, i.e., the clock parameters of distributed nodes, geometry-based applications such as positioning and sensing further require knowledge of the nodes' \textit{spatial states}
~\cite{Lu23AnchorState,Chen24nearField}.
The temporal states typically comprise clock time offsets (TOs), carrier frequency offsets (CFOs), and carrier phase offsets (CPOs).
The spatial states typically comprise node position, orientation, and velocity in scenarios involving mobility.

\subsection{Objective: Spatiotemporal Synchronization}
IEEE SA P3343 aims to establish algorithms and performance bounds for spatiotemporal synchronization, i.e., the joint inference of spatial and temporal states in networks of distributed apertures.
The term \emph{aperture} here implies that nodes are equipped with sensor arrays, giving them the ability to resolve incident signals in the angular domain, which comes at the cost of the unknown node orientation entering its spatial state. 
The most demanding, yet most rewarding aspect of this joint estimation problem is to establish phase-coherent operation of distributed nodes.
This implies challenging real-time estimation of unknown clock phases (online phase calibration),
but holds large performance gains
~\cite[Fig.\,4]{Fascista25JSTSP}.
If solved successfully, full spatiotemporal synchronization will transform a network of distributed apertures into a very large jointly coherent synthetic aperture, enabling high-rate communications, precise positioning and sensing, and efficient WPT.

\newcommand{\state}[1]{\bm{\theta}_{\scriptscriptstyle #1}}
\newcommand{\stateSpatial}[1]{\bm{\theta}_{\scriptscriptstyle #1}^{\text{\tiny{s}}}}
\newcommand{\stateTemporal}[1]{\bm{\theta}_{\scriptscriptstyle #1}^{\text{\tiny{t}}}}
\newcommand{\observation}[2]{\bm{z}_{\scriptscriptstyle #1,#2}}   
\newcommand{\factor}[2]{f_{\scriptscriptstyle #1#2}}   
\newcommand{\APpair}[2]{(\!#1,#2\!)}            
\newcommand{\setPairs}{\mathcal{P}}
\newcommand{\setApertures}[0]{\mathcal{J}}      
\newcommand{\realset}[2]{ \mathbb{R}^{#1 \times #2}  }
\newcommand{\realsetone}[1]{\mathbb{R}^{#1}}
\newcommand{\complexset}[2]{ \mathbb{C}^{#1 \times #2}  }
\newcommand{\complexsetone}[1]{ \mathbb{C}^{#1}  }
\newcommand{\stateSpace}[0]{\mathcal{S}_{\scriptscriptstyle\!\bm{\theta}}}                            
\newcommand{\herm}{\mathsf{H}}
\newcommand{\hermit}{\mathsf{H}}
\newcommand{\transp}{\mathsf{T}}
\newcommand{\trp}{\mathsf{T}}
\newcommand{\setAgents}[0]{\mathcal{M}}         
\newcommand{\setAnchors}[0]{\mathcal{A}}        
\newcommand{\setNeighbors}[1]{\mathcal{M}_{\scriptscriptstyle #1}}  
\newcommand{\Nchannel}[0]{N}                    
\newcommand{\naperture}[0]{j}                   
\newcommand{\nagent}[0]{m}                      
\newcommand{\nanchor}[0]{a}                     
\newcommand{\pos}[1]{ \bm{p}_{\scriptscriptstyle #1} }                                  
\newcommand{\posEstimate}[1]{ \hat{\bm{p}}_{\scriptscriptstyle #1} }                    
\newcommand{\etarot}[1]{\bm{\eta}^{\text{\tiny{o}}}_{\scriptscriptstyle#1}}
\newcommand{\stateHat}[1]{\hat{\bm{\theta}}_{\scriptscriptstyle #1}}
\newcommand{\covarianceEstimate}[1]{\widehat{\bm{C}}_{\scriptscriptstyle#1}}                
\newcommand{\stateHatMMSE}[1]{\hat{\bm{\theta}}^{\text{\tiny MMSE}}_{\scriptscriptstyle #1}}
\newcommand{\toffset}[1]{\epsilon_{#1}}                                                  

\begin{figure}[t]%
    \centering%
    \includegraphics[width = 252.0pt
    ]{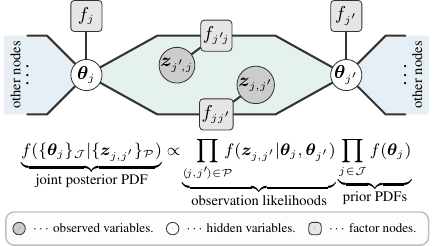}%
    \vspace{-0.2cm}\caption{Factor graph for two example nodes $j$ and $j'$ representing the joint posterior PDF of the states $\{\state{j}\}_{\setApertures}$. 
    We abbreviate $\factor{j}{j'}:=f\big(\observation{j}{j'}|\state{j},\state{j'}\big)$ and $\factor{j}{}:=f\big(\state{j}\big)$.}%
    \label{fig:factor-graph}%
\end{figure}%


\subsection{Standard Development Directions}

Let's consider a set $\setApertures:= \{1 \dots J\} = \setAgents \cup \setAnchors$ of spatially distributed apertures comprising a set of agents $\setAgents$ with unknown spatiotemporal parameters $\{\state{j}\}_{\scriptscriptstyle j \in \setAgents}$ and a set of anchors $\setAnchors$ with known spatiotemporal parameters $\{\state{j}\}_{\scriptscriptstyle j \in \setAnchors}$.
The spatiotemporal parameters $\state{j} :=\left[{\stateSpatial{j}}^\trp, \ {\stateTemporal{j}}^\trp \right]^\trp$ comprise the spatial state $\stateSpatial{j}$ (position, orientation, and velocity) and temporal state $\stateTemporal{j}$ (TO, CFO, and CPO) of each aperture $j \in \setApertures$.
Assuming independent bidirectional measurements $\observation{j}{j'}$ between all $|\setPairs|\!=\!J(J\!-\!1)$ ordered aperture pairs $\APpair{j}{j'} \!\in\! \setPairs \!:=\! \big\{\APpair{j}{j'} \!\in\! \setApertures^2 | j\!\neq\! j'\big\}$, the \textit{joint} likelihood $f\big(\{\observation{j}{j'}\}_{\scriptscriptstyle \setPairs}|\{\state{j}\}_{\scriptscriptstyle \setApertures}\big) = \prod_{\APpair{j}{j'}\in\setPairs}f\big(\observation{j}{j'}|\state{j},\state{j'}\big)$ factorizes into a product of local observation likelihoods between all aperture pairs~\cite{Meyer13particleBasedBP}.
Likewise, assuming a priori independent spatiotemporal aperture states, the \textit{joint} prior probability density function (PDF) factorizes as $f(\{\state{j}\}_{\scriptscriptstyle \setApertures})=\!\prod_{\scriptscriptstyle j\in\setApertures}\!f(\state{j})$.
The \textit{joint} posterior PDF (up to a proportionality constant) of all spatiotemporal aperture states conditional on all observations is represented by the factor graph~\cite{Loeliger04IntroFG} in Fig.\,\ref{fig:factor-graph} which---by exploiting the conditional independence structure that the graph makes explicit---can be shown to factorize as $f(\{\state{j}\}_{\scriptscriptstyle\setApertures}|\{\observation{j}{j'}\}_{\scriptscriptstyle\setPairs})%
\!\!\propto\!%
\!\prod_{\scriptscriptstyle\APpair{j}{j'}\in\setPairs} \!f(\observation{j}{j'}|\state{j},\state{j'})%
\!\prod_{\scriptscriptstyle j\in\setApertures}\!f(\state{j})$ into a product of the bidirectional observation likelihoods between all aperture pairs $\APpair{j}{j'}\in \setPairs$ with prior PDFs $f(\state{j})$ of the spatiotemporal states of all apertures $j \in \setApertures$. 
The factor graph represents the joint statistical model of the spatiotemporal synchronization problem showing two example nodes $\APpair{j}{j'}$ while the ``other nodes'' $\setPairs \setminus \APpair{j}{j'}$ are not depicted for brevity.
The maximum a posteriori (MAP) or the minimum mean square error (MMSE) estimates are computed as the maximum or expectation, respectively, under the \textit{marginal} posterior PDFs $f(\state{j}|\{\observation{j}{j'}\}_{\scriptscriptstyle\setPairs})$.
A popular technique to efficiently obtain the marginal posterior PDFs and avoid a possibly prohibitive direct marginalization is the sum-product algorithm (SPA)~\cite[Sec.\,8.4.4]{Bishop}, which would result in \textit{exact} inference in tree-structured factor graphs. 
The spatiotemporal synchronization problem, however, falls into the realm of cooperative localization and synchronization, where the graph structure usually has cycles. 
Loopy belief propagation (BP) applies SPA message passing in an iterative manner, providing \textit{approximate} inference in cyclic factor graphs~\cite{Frey97loopyBP}. 
Computation of beliefs (unnormalized marginal posterior PDFs) still involves marginalization integrals\footnote{In messages from factor nodes to variable nodes~\cite[eq.\,8.66]{Bishop}.}.
Sigma point BP~\cite{Meyer14sigmaPointBP,Masiero25sigmaPointSLAM} or efficient particle-based BP~\cite{Meyer13particleBasedBP,Deutschmann25CISA,Wielandner21PosOrient,Wielandner9Dcooperative,Leitinger24mpSLAMmapFusion} can circumvent explicit integration~\cite{Meyer15diss} and provide attractive methodological solutions.
In practical problems, further parameters enter the statistical model.
Spatiotemporal synchronization with loopy BP can either be implemented in (i) a two-step fashion~\cite{Meyer13particleBasedBP}, which requires the use of a preceding channel estimator that extracts angle, delay, and Doppler estimates, or in (ii) a direct fashion~\cite{Deutschmann25CISA}, working directly on noisy channel observations. 
Nuisance parameters such as amplitudes or noise power of channel observations can be treated by marginalization~\cite{Wielandner21NuisanceParameters} or concentration~\cite{Deutschmann25CISA,Deutschmann25OJSP}.
In channels with specular multipath components (SMCs), spatiotemporal synchronization is complicated by the data association problem~\cite{Mendrzik19SMCPosOrient2D,Leitinger19BP_SLAM} for which efficient implementations exist~\cite{Meyer15MTT}.

IEEE SA P3343 aims to define cooperative‑inference algorithms and their performance bounds for spatiotemporal synchronization in networks of distributed apertures. 
The methods reviewed above enable robust D‑MIMO, accurate multistatic‑ISAC, and efficient WPT ecosystems for 6G and beyond.

%

%
%

%
%

\section*{Conclusion}
The standardization process for emerging technologies is a key enabler for continued maturation and eventual commercial success.  The P3343 and P3383 working groups of the SPS-SASC are developing standards for ISAC performance metrics and for the spatio-temporal synchronization of distributed apertures.  The final standards proposed by the working groups will also benefit the academic community by defining best practices and the current state-of-the-art.  ISAC performance metrics are particularly important to the continued evolution of 6G wireless communications and standardized techniques for synchronizing distributed apertures in both time and space are essential to the operation of multi-static sensing platforms such as those employed in radar.
%
\bibliographystyle{IEEEtran}
\bibliography{refs}







%
%
%
%
\section*{Biographies}
\label{sec:bio}
%
%
%
%

%


\begin{IEEEbiography}[{\includegraphics[width=1in,height=1.25in,clip,keepaspectratio]{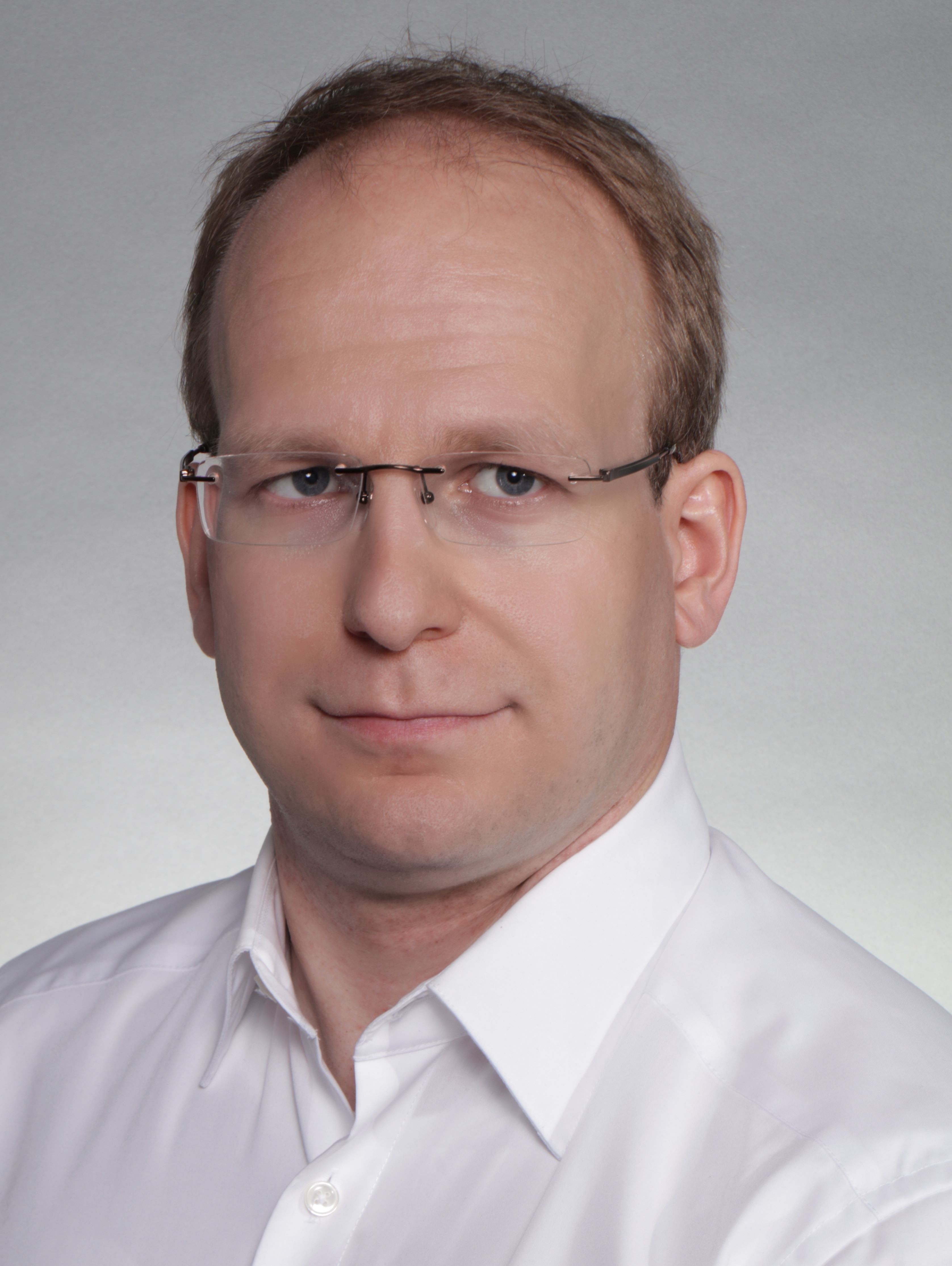}}]{\noindent Andreas Bathelt}   received the M.Sc. degree in electrical engineering from the Mannheim University of Applied Sciences, Germany in 2012 and the Dr.-Ing. in the field of Control Theory/System Identification from the University of Duisburg/Essen, Germany in 2018. He was a Research Assistant at the Faculty of Process Engineering, Energy and Mechanical Systems of the TH Köln – University of Applied Sciences from 2012 to 2017. Here, he was working in the field of Control Theory and System Identification. He joined the Fraunhofer Institute for High Frequency Physics and Radar Techniques FHR  as a Researcher in 2017. Within in the general scope of Radar, his research interests are in Integrated Sensing and Communications (ISAC), where the focus is on methods and signaling for sensing with minimal spectral usage, and Time Synchronization with a focus on consensus-based time synchronization in multi-agent systems. He is a currently the vice chair of the IEEE Standards Association P3383 Working Group.
\end{IEEEbiography}

\begin{IEEEbiography}[{\includegraphics[width=1in,height=1.25in,clip,keepaspectratio]{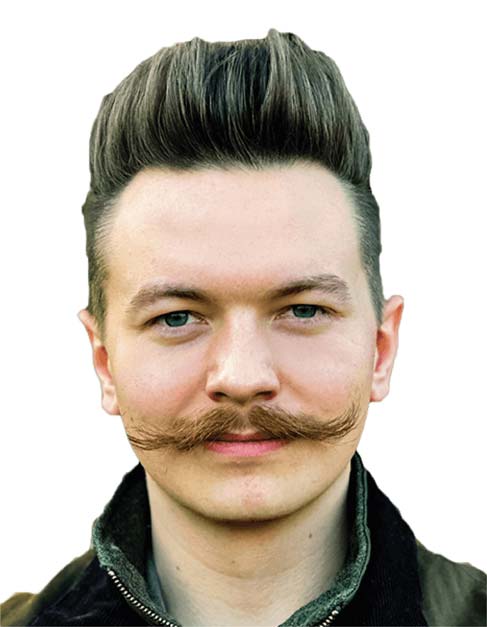}}]{\noindent Benjamin J.\,B. Deutschmann} (~\IEEEmembership{Graduate~Student~Member,~IEEE}) received the B.Sc.\ (2017) and Dipl.-Ing.\ (2021) degrees in electrical engineering from Graz University of Technology, Graz, Austria. 
From 2017 to 2021 he was with the Institute of Microwave and Photonic Engineering, investigating loosely coupled inductive links for wireless power transfer and RF impedance-measurement systems. 
He is currently pursuing the Ph.D.\ degree at the Institute of Communication Networks and Satellite Communications, where his research interests include wireless channel modeling, array signal processing, positioning and mapping, and wireless power transfer with distributed and large-aperture antenna arrays.
He is a member of the IEEE Signal Processing Society Synthetic Aperture Standards Committee and chairs the IEEE Standards Association P3343 Working Group.
\end{IEEEbiography}

\begin{IEEEbiography}[{\includegraphics[width=1in,height=1.25in,clip,keepaspectratio]{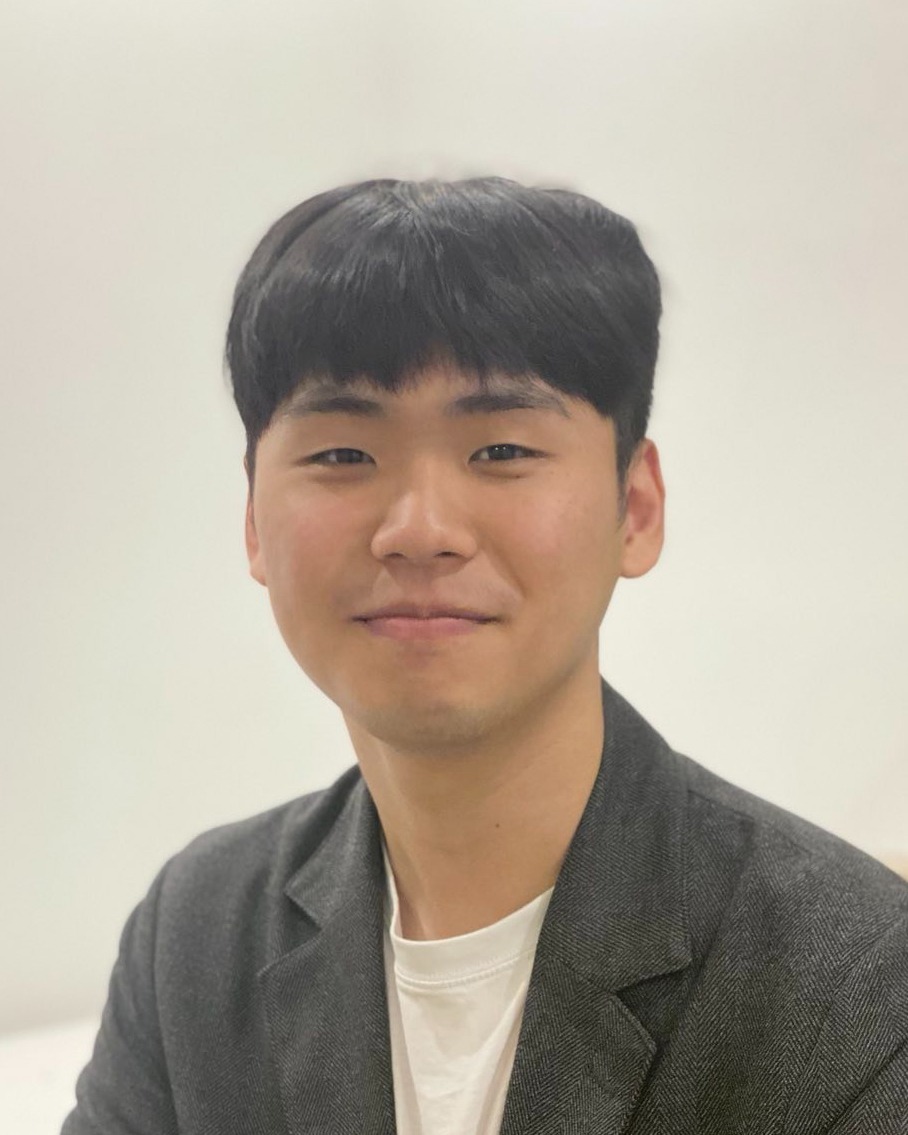}}]{\noindent Hyeon Seok Rou} (~\IEEEmembership{Member,~IEEE}) received the B.Sc. degree in electrical and computer engineering and the Ph.D. degree in electrical engineering from Constructor University (formerly Jacobs University Bremen), Bremen, Germany, in 2021 and 2024, respectively. Since 2025, he is currently a postdoctoral fellow and a lecturer with the School of Computer Science and Engineering, Constructor University Bremen. He was a Visiting Researcher with the Intelligent Communications Laboratory (ICL!), Korea Advanced Institute of Science and Technology (KAIST), in 2023. He has received Korea Institute of Science and Technology Europe Research Scholarship Award from The Korean Scientists and Engineers Association in the FRG (Verein Koreanischer Naturwissenschaftler und Ingenieure in der BRD e.V., VeKNI e.V.) in 2022, and is a representative at the Young Professionals Forum (YPF) of the 3rd World Congress of Korean Scientists and Engineers, South Korea, 2025. His research interests include integrated sensing and communications (ISAC), signal processing in doubly-dispersive channels, high-mobility communication systems, next-generation metasurfaces, B5G/6G V2X technologies, and quantum computing.
\end{IEEEbiography}

\begin{IEEEbiography}[{\includegraphics[width=1in,height=1.25in,clip,keepaspectratio]{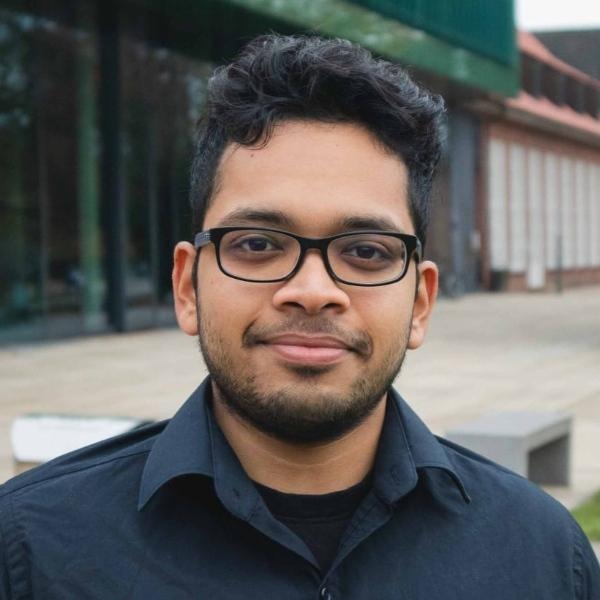}}]{Kuranage Roche Rayan Ranasinghe} (~\IEEEmembership{Graduate Student Member,~IEEE}) received the B.Sc. degree in electrical and computer engineering and the B.Sc. degree in robotics and intelligent systems (with a minor in physics) from Constructor University (formerly Jacobs University Bremen) in 2023 and 2024, respectively, where he is currently pursuing the Ph.D. degree in electrical engineering. Within the fields of wireless communications and signal processing, his research interests encompass integrated sensing, communications and computing (ISCC), compressed sensing, Bayesian inference, next-generation metasurface technologies, and optimization theory. He received the Best Paper Award at the International Conference on Computing, Networking and Communications (ICNC) in 2025.
\end{IEEEbiography}

\vfill

\begin{IEEEbiography}[{\includegraphics[width=1in,height=1.25in,clip,keepaspectratio]{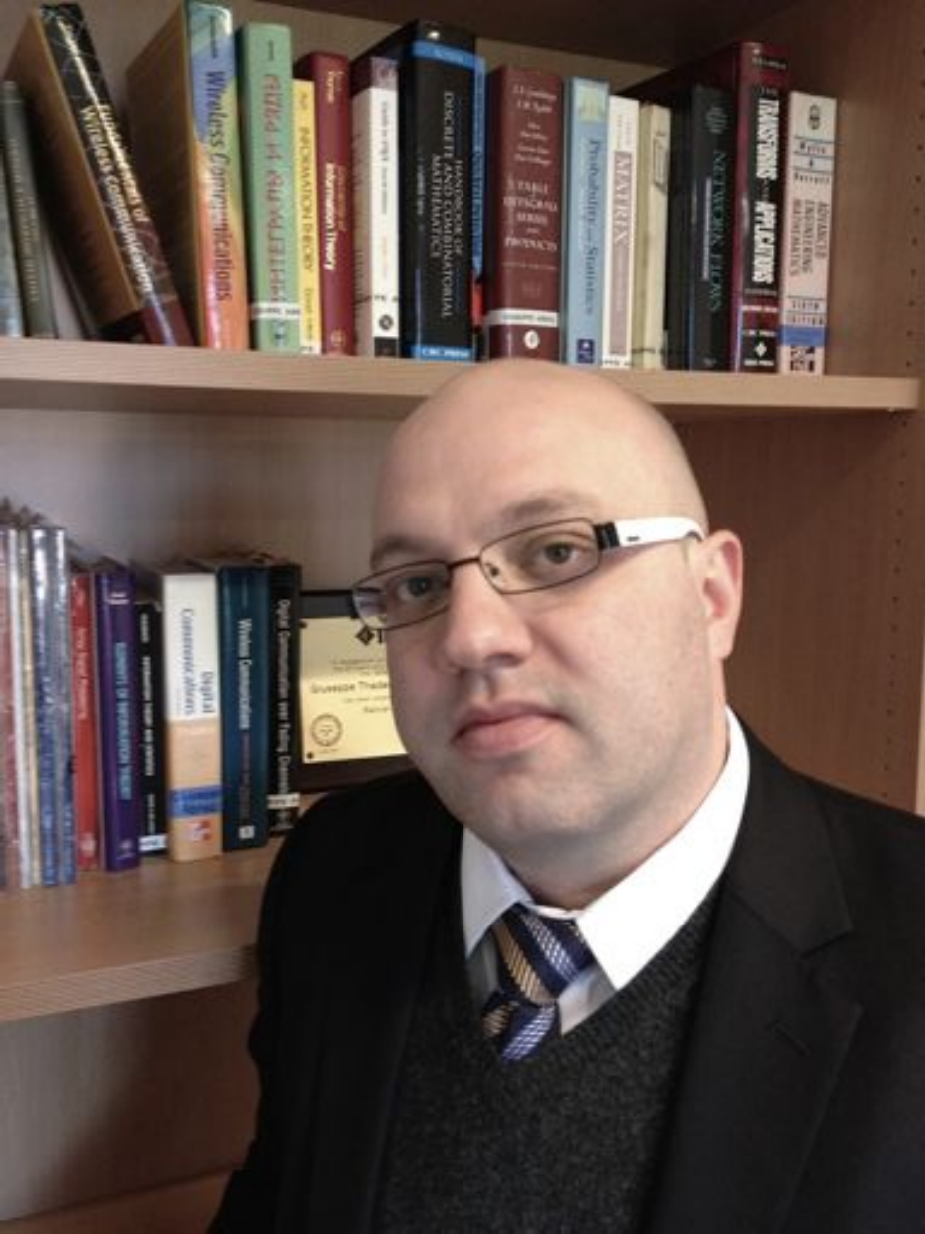}}]{Giuseppe Thadeu Freitas de Abreu} (~\IEEEmembership{Senior Member,~IEEE}) received the B.Eng. degree in electrical engineering and the specialization Latu Sensu degree in telecommunications engineering from the Universidade Federal da Bahia (UFBa), Salvador, Bahia, Brazil in 1996 and 1997, respectively, and the M.Eng. and D.Eng. degrees in physics, electrical, and computer engineering from Yokohama National University, Japan, in March 2001 and March 2004, respectively. He was a postdoctoral fellow and later an adjunct professor (docent) in statistical signal processing and communications theory at the Department of Electrical and Information Engineering, University of Oulu, Finland from 2004 to 2006 and from 2006 to 2011, respectively. Since 2011, he has been a professor of electrical engineering at Constructor University (formerly known as Jacobs University), Bremen, Germany. From April 2015 to August 2018, he also simultaneously held a full professorship at the Department of Computer and Electrical Engineering, Ritsumeikan
University, Japan.
His research interests include communications and signal processing, communications theory, estimation theory, statistical modeling, wireless localization, cognitive radio, wireless security, MIMO
systems, ultrawideband and millimeter wave communications, full-duplex and cognitive radio, compressive sensing, energy harvesting networks, random networks, connected vehicles networks, electro-magnetic signal processing, quantum computing for signal processing, metasurfaces for wireless systems and many other topics.
Prof. Abreu has received various awards and prestigious fellowships, including the Uenohara Award from Tokyo University in 2000, the prestigious JSPS, Heiwa Nakajima, and NICT (twice)
fellowships in 2010, 2013, 2015, and 2018, as well as Best Paper award at several international
conferences, and of the Best Paper award by the Japanese Chapter of the IEEE Signal Processing Society in 2023.
He served as an associate editor for the IEEE Transactions on Wireless Communications from 2009 to
2014 and the IEEE Transactions on Communications from 2014 to 2017;
as an executive editor for IEEE Transactions on Wireless Communications
from 2017 to 2021, and as an editor IEEE Communications Letters from 2021 to 2023.
He is currently serving as an editor to the IEEE Signal
Processing Letters and to the IEEE Open Journal of the Communications Society.
\end{IEEEbiography}

\begin{IEEEbiography}[{\includegraphics[width=1in,height=1.25in,clip,keepaspectratio]{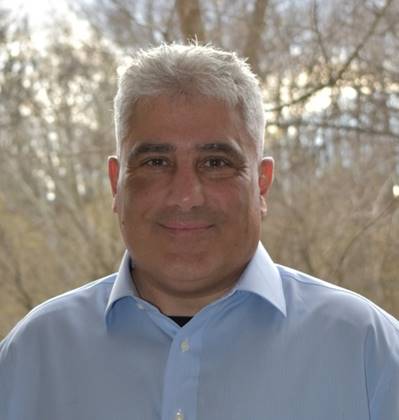}}]{Peter Vouras} (~\IEEEmembership{Senior Member,~IEEE}) was the founding Chair of the IEEE Signal Processing Society Synthetic Aperture Technical Working Group (SA-TWG) and is the current Chair of the IEEE Signal Processing Society Synthetic Aperture Standards Committee (SASC).  The SASC works to develop technical standards that define best practices for image formation using synthetic apertures in radar, channel sounding, radiometry, magnetic resonance imaging, and sonar.  Mr. Vouras obtained a M.S.E degree in Electrical and Computer Engineering from the Johns Hopkins University, Homewood Campus, Baltimore, MD in 2001, a B.S. degree in Electrical and Computer Engineering from George Mason University, and B.A. degrees in Economics and Foreign Affairs from the University of Virginia.  He is currently with the U.S. Department of Defense and was previously with the Wireless Networks Division of the Communications Technology Laboratory at the National Institute of Standards and Technology in Gaithersburg, MD (2016 to 2022).   From 1996 to 2016, Mr. Vouras was with the Radar Division at the Naval Research Laboratory in Washington, D.C.  His research interests include computational imaging, remote sensing, and inverse problems.
\end{IEEEbiography}

\vfill

\end{document}